# Highly Volcanic Exoplanets, Lava Worlds, and Magma Ocean Worlds: An Emerging Class of Dynamic Exoplanets of Significant Scientific Priority

A white paper submitted in response to the National Academy of Sciences 2018 Exoplanet Science Strategy solicitation, from the NASA Sellers Exoplanet Environments Collaboration (SEEC) of the Goddard Space Flight Center

Wade G. Henning[1,2]*, Joseph P. Renaud[3], Prabal Saxena[1], Patrick L. Whelley[1,2], Avi M. Mandell[1], Soko Matsumura[4], Lori S. Glaze[1], Terry A. Hurford[1], Timothy A. Livengood[1,2], Christopher W. Hamilton[5], Michael Efroimsky[6], Valeri V. Makarov[6], Ciprian T. Berghea[6], Scott D. Guzewich[1], Kostas Tsigaridis[7, 12], Giada N. Arney[1], Daniel R. Cremons[1], Stephen R. Kane[8], Jacob E. Bleacher[1], Ravi K. Kopparapu[1], Erika Kohler[1], Yuni Lee[1], Andrew Rushby[9], Weijia Kuang[1], Rory Barnes[10], Jacob A. Richardson[1], Peter Driscoll[11], Nicholas C. Schmerr[2], Anthony D. Del Genio[12], Ashley Gerard Davies[13], Lisa Kaltenegger[14], Linda Elkins-Tanton[15], Yuka Fujii[16], Laura Schaefer[15], Sukrit Ranjan[17], Elisa Quintana[1], Thomas S. Barclay[1,2], Keiko Hamano[16], Noah E. Petro[1], Jordan D. Kendall[1,2], Eric D. Lopez[1], Dimitar D. Sasselov[18]

**Author Affiliations:**
1. NASA Goddard Space Flight Center, 2. University of Maryland, 3. George Mason University, 4. University of Dundee, 5. University of Arizona, 6. United States Naval Observatory, 7. Columbia University, 8. University of California, Riverside, 9. NASA Ames Research Center, 10. University of Washington, 11. Carnegie Institution for Science, 12. NASA Goddard Institute for Space Studies, 13. Jet Propulsion Laboratory - California Institute of Technology, 14. Carl Sagan Institute, Cornell University, 15. Arizona State University, 16. Earth-Life Science Institute, Tokyo Institute of Technology, 17. Massachusetts Institute of Technology, 18. Harvard University

* Corresponding Author: 301-614-5649  wade.g.henning@nasa.gov

**Acronyms:**

| | |
|---|---|
| NExSS | Nexus for Exoplanetary System Science |
| HST | Hubble Space Telescope |
| JWST | James Webb Space Telescope |
| OST | Origins Space Telescope |
| HabEx | Habitable Exoplanet Imaging Mission |
| LUVOIR | Large Ultraviolet Optical and Infrared Surveyor |
| TESS | Transiting Exoplanet Survey Satellite |
| PLATO | Planetary Transits and Oscillations of Stars |
| MMR | Mean Motion Resonance |
| SNR | Signal to Noise Ratio |
| GCM | General Circulation Model |
| HZ | Habitable Zone |
| STEM | Science Technology Engineering and Mathematics |

**Introduction:**
   Highly volcanic exoplanets, which can be variously characterized as 'lava worlds', 'magma ocean worlds', or 'super-Ios' are high priority targets for investigation. The term 'lava world' may refer to any planet with extensive surface lava lakes, while the term 'magma ocean world' refers to planets with global or hemispherical magma oceans at their surface. 'Highly volcanic planets' may simply have large, or large numbers of, active explosive or extrusive volcanoes of any form. They are plausibly highly diverse, with magmatic processes across a wide range of compositions, temperatures, activity rates, volcanic eruption styles, and background gravitational force magnitudes. Worlds in all these classes are likely to be the most characterizable rocky exoplanets in the near future due to observational advantages that stem from their preferential occurrence in short orbital periods and their bright day-side flux in the infrared. Transit techniques should enable a level of characterization of these worlds analogous to hot Jupiters. Understanding processes on highly volcanic worlds is critical to interpret imminent observations. The physical states of these worlds are likely to inform not just geodynamic processes, but also planet formation, and phenomena crucial to habitability. Volcanic and magmatic activity uniquely allows chemical investigation of otherwise spectroscopically inaccessible interior compositions. These worlds will be vital to assess the degree to which planetary interior element abundances compare to their stellar hosts. We suggest that highly volcanic worlds may become second only to habitable worlds in terms of both scientific and public long-term interest.

**Background and Science Opportunities:**
   Three energy sources generally work together to power sustained activity on highly volcanic worlds: (1) radionuclides, (2) insolation, and (3) tidal heating. Giant impacts provide transient heat, as the presumptive Moon-forming impact temporarily converted Earth to a magma ocean.

   **Radionuclides**, for a young Earth analog, can lead to volcanism rates ~10× the modern Earth rate. If formed closer to the center of the galaxy, near a recent supernova, or near a neutron star merger, young planet radiogenic rates may be far higher. Many lava world candidates are in the super-Earth category (masses ~1–10× that of the Earth's), as a larger mantle mass leads to more radiogenic material, fueling intense volcanic activity by a lower surface-to-volume ratio.

   **Insolation** does not drive volcanism alone, but can set extremely high surface temperature boundary conditions (1500–3000 K) that exceed the melting point of many mineral species. Such extreme radiative environments may produce worlds with localized to hemispheric surface magma oceans [1, 2, 3]. Mass transport from these oceans may produce evolved magma compositions and variegated surfaces where deposition is occurring [3]. Not only are such processes likely to influence observational signatures (e.g., through phase curve variations), but the ability to detect their manifestations may help constrain interior processes, volatile reservoirs and a host of otherwise inaccessible planetary properties. Insolation may establish lava worlds and magma worlds without excess internal heat. Additionally, insolation controls the existence, structure, and dynamics of atmospheres on such planets. For observations, magma–atmosphere interactions will be critical. It is an open question as to what magma ocean world conditions, if any, might allow for the pressure build-up required for explosive-style eruptions.

   **Tidal Heating** has the potential to generate internal heat rates millions of times the output of the modern Earth [4]. The exact upper limits to such heating are uncertain, and will be driven by the interplay between mantle partial melting, heat-pipe style magmatic escape from deep-mantle sources, and issues concerning magma buoyancy and processing according to melting (solidus) temperatures. Tidal heat of such magnitudes may be short-lived, such as terrestrial worlds scattered temporarily into high eccentricity orbits [5], or driven into non-synchronous spin states by scattering or collisions. Alternately, intense heating may be sustained for billions of years in quasi-equilibrium with mantle convection or advective cooling, if a planet is in MMR with a large perturber, akin to Io in our Solar System. This is particularly important given the number of compact near-MMR multi-planet systems already detected (including TRAPPIST-1).

**Example Exoworlds**: Candidate worlds include those that are so close to their host stars that they are nearly disintegrating [6], all the way to outer planets/moons whose volcanism may be tidally driven. Top candidates include: 55 Cnc e, TRAPPIST-1 b and c, COROT-7b, Kepler-10b, WASP-47e, Kepler-78b, GJ 581e and GJ 1132b. Systems such as TRAPPIST-1, Tau Ceti, and Kepler-90, suggest compact multibody terrestrial systems in Laplace-like MMR chains or near-resonance are not uncommon. Such conditions are ideal for significant sustained tidal heating and magmatic activity, which may thus be occurring at worlds only a few lightyears away.

**Observational Opportunities:** The effects of large volcanic eruptions will be observable on exoplanets with JWST [7, 8]. Given the biases inherent in the most prolific exoplanet detection methods, highly volcanic planets are going to be some of the first characterizable rocky worlds. Even relatively low SNR observations of 55 Cnc e have already yielded fascinating results that require unique physical mechanisms as an explanation. Eclipse and transit variability, poor day–night heat redistribution, a relatively hot nightside, and tentative detections of a sodium atmosphere have all been observed for 55 Cnc e [9, 10, 11], and underpin the relatively unknown surface environments on these worlds. Current and upcoming surveys such as TESS, PLATO, and ground based surveys promise to find even more of these worlds, while JWST and next generation ground based observatories will revolutionize our ability to investigate them. Volcanic activity offers numerous pathways for observations, including: phase curve variations [12, 13], transmission spectroscopy of stratospheric plumes and volcanically-generated secondary atmospheres [7] (which could provide otherwise unobservable measurements of a planet's geochemistry and composition), direct IR observations of magma ocean surfaces or lava lakes [14, 15], or dynamical inference from eccentricities or rotation rates [5].

**Habitability:** Highly volcanic worlds and the habitable-worlds class overlap in critical ways. Volcanoes are well-established as essential for high-biomass surface life by refreshing surface volatile elements and redox potentials generally, via the gradual outgassing of a mantle. Volcanic outgassing modulates surface temperatures via sulfur compound release, aerosol production, ash production, mantle degassing of water, and as the central component of carbon-cycle climate feedback. Indeed, the only known inhabited world is also volcanically active, where subaerial volcanoes alone provide tens of teragrams of $CO_2$ to the surface annually [16]. This suggests a new volcanic rate-dependent 'Goldilocks Zone', from minimal magma intrusion in the shallow subsurface that can host low-biomass extremophile/deep biosphere life, to ideal redox states for high biomass activity, up to volcanically prolific worlds that create a greenhouse but without full mass extinction. TESS, JWST, and beyond will begin to probe this Goldilocks Zone from the 'Venus/Hot' end-member toward the 'Mars/Cold' end-member. Extreme submarine volcanism on inundated planets may severely alter ocean chemistry, yet be challenging to detect. Such habitability alterations may also be mediated by volcanism arising from tidal heat [17, 18].

**Relevance to All Silicate Terrestrial Worlds:** Supervolcanic worlds typically cool in time as radionuclides decay, and may evolve into habitable worlds in later stages. Observational constraints that can be found *only* from observing magma ocean worlds are in part, a form of time machine: allowing a window into studying the magma ocean phase with which the Earth, Mars, Moon, and perhaps a majority of all terrestrial exoplanets passed through after initial accretion and mantle outgassing. Efforts to compare volcanism across planets in our Solar

System suggest that radiogenic magma production rates scale to planetary radius [19]. Observations of volcanic exoworlds would enable testing this hypothesis with a larger number of objects, which will improve our expectations for volcanic histories of all rocky worlds.

**Dynamical Role:** Supervolcanic activity may also directly support the survival of terrestrial worlds against orbital scattering and loss into host stars or to system ejection, via the action of extreme tides/rapid-eccentricity-damping preventing orbit crossings with perturbers [5]. Supervolcanic activity also plays a keystone role in spin synchronization, as during despin planets will often experience high tidal friction, interiors will be altered, and this may lead planets to: (a) non 1:1 spin-orbit resonances [20, 21], or (b) pseudo-synchronous rotation rates a few percent faster than synchronous [22]. A planet in high spin-orbit resonance can also warm up and spontaneously depart resonance when viscosity drops below a certain threshold, eventually ending up in the less dissipative 1:1 state. Mass redistribution will also alter spin dynamics and may lead to extreme true-polar wander, upsetting long-term dayside/nightside dichotomies even on 1:1 locked worlds. All such events require testing within exoplanet climate (GCM) research.

**Io and Exo-Io/super-Io Worlds:** Io provides a local archetype of a diverse category of related silicate worlds with intense tidally-driven volcanism [23] and is a high priority for focused investigation [24, 25]. Future surveys of Io-analogs can help test common assumptions regarding how stellar abundances translate to planetary compositions. An Io-analog world adjacent to a habitable world will be especially valuable, allowing heavier element abundances for the system to be investigated better than at minimally-volcanic worlds in isolation. Tidal heating is expected to be particularly important for planets orbiting M-dwarfs where the radiative HZ overlaps the predicted tidal zone where tidal heating may dominate the internal heat budget [26]. Theoretical development is needed to further couple tidal heating, internal structure, thermal/orbital evolution, and investigate their role in a range of system architectures.

**Venus and Exo-Venus Planets:** Volcanism, could, in principle, be remotely detected on Venus and exo-Venus planets, a subset of which could be highly volcanic. Remote searches for volcanism on Venus below the clouds are enabled by "near-infrared spectral windows" on the planet's nightside [27] between 1–2.5 μm that sense surface and lower atmosphere thermal emission [28]. Venus Express observed evidence for volcanism on Venus via four temporally variable hotspots at volcanoes Ozza Mons and Maat Mons in the Ganis Chasma rift zone [29]. Above the clouds (>50 km altitude), occasional rapid increases in $SO_2$ have been suggestive of infrequent large eruptions [30, 31, 32, 33]. On an exo-Venus, evidence for volcanism could come via nightside observations that measure sub-cloud radiation and reveal time-variable emissions of volcanic gases. Observing the nightside of a spatially unresolved exoplanet requires high SNR measurements at crescent phase so reflected dayside radiation can be removed. On the dayside, $SO_2$ can be measured at UV wavelengths to search for time-variable volcanic emissions. Precise transit spectroscopy, which could measure radiation above the cloud deck of an exo-Venus, may be able to detect periodic volcanogenic gases. For true exo-Venus analogs, such gas species will likely need to be detectable above an optically thick lower atmosphere. Volcanic plumes are less buoyant in hot and dense atmospheres, and thus would be expected to reach high (and detectable) altitudes less frequently than on planets with colder, thinner atmospheres [32, 34].

Below, we highlight topics outlined by the National Academy of Sciences 2018 Exoplanet Science Strategy solicitation:

– **Identify areas of significant scientific progress since publication of the New Worlds New Horizons Decadal Survey.**
• Temporal/spatial variations of 55 Cnc e have been detected [9] and may be geologic in origin.

- Kepler data results now imply a large population of terrestrial worlds exist. All new Kepler rocky bodies, down to approximately Mars masses, open an era of terrestrial world study.
- Detection of systems with complex, multi-body dynamics, such as TRAPPIST-1, Kepler-90, Tau Ceti, & 55 Cnc (with some components at or near MMRs), favors ongoing tidal activity.

– **Identify exoplanet science areas where significant progress will likely be made with current and upcoming observational facilities.**

- TESS, JWST, and any LUVOIR/HabEx/OST flagship will locate and enable characterization of nearby highly volcanic worlds [14]. In particular, the NIR and MIR spectroscopic capabilities of JWST provide an excellent tool for examining magma ocean thermal flux.
- Many TESS worlds may be volcanic but not in a HZ. TESS will focus on nearby M-dwarf stars, with less extreme surface insolation, where tidal/internal heat has better signal.
- Detection of exomoons might begin a long process to constrain Io-analog occurrence rates.
- Exo-Venus atmospheres (perhaps Kepler-1649b [35]) could be characterized within a decade.
- JWST can detect explosive large stratospheric eruptions (Pinatubo-class) out to ~25 lightyears [7], but is limited by observing time to catch rare events. For Earth, the mixing & residence time of volcanogenic $SO_2$ prior conversion to $H_2SO_4$ aerosol haze is ~3 months.
- Non-explosive outgassing and vapor partial pressures [36] for volcanic/lava worlds may increase atmospheric detectability, and not require catching rare events.

– **Identify exoplanet science areas and key questions that will likely remain after these current and planned missions are completed.**

- *The Io Volcano Observer* [24]*, or related efforts to monitor Io, should be considered a high priority*, as the best path to address an entire class of scientifically critical active exoplanets.
- Modeling of Venus-analog environments is required to constrain expected high atmosphere $SO_2$ signatures of volcanic eruptions, and determine what levels above various backgrounds would be required for UV or IR detection by future missions on exo-Venus planets.
- Detection of Venus analogs may reach a characterization bottleneck at haze layers until direct emission spectra are obtainable. Studies must evaluate what advances/parameters will enable measurements through opaque atmospheres (e.g., wavelengths of atmospheric windows for diverse haze compositions, integration times and repetition rates required, and atmospheric penetration depths required for eruption columns of differing buoyancy).
- Large explosive eruptions on Earth-analog worlds may not be seen by JWST/future flagship primary missions, unless monitoring cadences are found feasible by observing panels.

– **Identify key observational, technological, theoretical, and computational challenges for further progress understanding exoplanetary systems. Identify resources and timescales.**

- Wavebands for possible upcoming observatories should take into consideration IR/UV bands useful for volcanic/magmatic characterization (e.g., $SO_2$ lines, 1–2.5 μm haze windows).
- Exoplanet eruption detection is primarily constrained by observing time limits. Catching rare large eruptions on candidate planets will require: automated re-screening of candidate worlds suites at ~1, 2, or 4 (1/yr) rates. This is programmatically challenging, but the science payoff would be enormous. *Future buildup of a diverse infrastructure of ground and space based redundant systems will help to provide sufficient observing time. Refueling and servicing of space-based assets is imperative to enable future monitoring of exoplanets for temporal variations. Exoplanet science directly benefits from human cislunar presence by this path.*
- Investigating the spatial distribution of continents, oceans, and dry-land variations is feasible

[37] due to global rotation (spin-orbit tomography), but will require careful future application. Carbonatite lavas [38], if ever detected, might serve as a proxy for past water oceans.
- A strong theoretical effort is required for coupled interior/atmosphere/orbital/stellar modeling of highly volcanic worlds, with compositional diversity presenting a major challenge.
- Magma ocean world atmospheres [39] contain vapors of Fe, Mg, Si, Al, Ca, K, Na, and Ti [36]. High temperature furnaces are required to study chemistry in the 2000–3500 K range. *Support for planetary major equipment installation, upgrades, and maintenance, is vital for close-in hot exoplanet study*. Hydrodynamic mass loss of heavy elements is possible for close-in hot atmospheres, and may systematically alter magma ocean compositions [40].

– **Identify fruitful cross-disciplinary and public/private partnerships, topics and initiatives that will enable and accelerate progress in future areas of exoplanet inquiry.**
- Earth, Io, Mars, and Lunar volcanologists should be actively invited to study exovolcanology.
- Expand use of the NASA HQ public input model (pTA) [41] for exoplanet geoscience topics.
- Public interest in supervolcanic worlds and magma ocean worlds is intrinsically very high, due to their alien/exotic nature. *Public and private institutions may broaden public engagement and STEM learning by using supervolcanic worlds as cornerstone examples of exoplanet science* (in talks, websites and printed material) just as often as using habitable worlds. Observations and discoveries in this emerging planet class may have an outsized impact on public awareness of exoplanet science.

In summary, we emphasize that astonishing breakthroughs in this field are within reach, and the key to unlock them is not the exact flagships selected, but more so the maintenance of a robust exoplanet focus in the coming decades, akin to the 30-year roadmap laid out in [42]. With a program not impeded by repeated cancellations and restarts, the next decades will witness, among other breakthroughs, the creation of the new field of exoplanet volcanology.